# Self Consistent Expansion for the Molecular Beam Epitaxy Equation

Eytan Katzav

Raymond and Beverly Sackler Faculty of Exact Sciences

School of Physics and Astronomy, Tel Aviv University,

Ramat Aviv, Tel Aviv 69978, Israel

Abstract

Motivated by a controversy over the correct results derived from the dynamic renormalization group (DRG) analysis of the non linear molecular beam epitaxy (MBE) equation, a self-consistent expansion (SCE) for the non linear MBE theory is considered. The scaling exponents are obtained for spatially correlated noise of the general form $D(\vec{r}-\vec{r}',t-t') = 2D_0 |\vec{r}-\vec{r}'|^{2\rho-d} \delta(t-t')$. I find a lower critical dimension $d_c(\rho) = 4 + 2\rho$, above which the linear MBE solution appears. Below the lower critical dimension a $\rho$-dependent strong-coupling solution is found. These results help to resolve the controversy over the correct exponents that describe non linear MBE, using a reliable method that proved itself in the past by giving reasonable results for the strong coupling regime of the Kardar-Parisi-Zhang (KPZ) system (for $d > 1$), where DRG failed to do so.

The field of disorderly surface growth has received much attention during the last two decades. Special effort has been focused on relating discrete microscopic growth models with their corresponding continuum field theories [1]. The first continuum equation used to study the growth of interfaces by particle deposition was the Edwards-Wilkinson model (EW) [2] that describes the dynamics of the interface by a noise driven diffusion equation. This model actually describes the microscopic process known as random deposition (RD) with surface relaxation, and together they form a distinct universality class in growth phenomena. However, an extension to this model was needed because of the nonlinear character of many deposition processes, such as ballistic deposition (BD), solid-on-solid deposition (SOS) and Eden growth. The first extension of the EW equation to include nonlinear terms was proposed by Kardar, Parisi and Zhang (KPZ) [3], who suggested the addition of a nonlinear term proportional to the square of the height gradient. The success of the KPZ equation in describing deposition phenomena motivated many researchers to develop a continuum growth model relevant for the technologically important molecular beam epitaxy (MBE) process [4-10]. The physical mechanism that distinguishes MBE from previously discussed growth processes is the surface diffusion of the deposited particles. It is well known that in the temperature range of MBE growth, desorption of atoms and formation of overhangs and bulk defects is negligibly small. As a consequence the continuum model describing this processes must conserve the number of particles on the interface. The introduction of conservation laws into the growth equation forms new universality classes in surface phenomena. One of these classes is known as the Linear MBE equation (or the Mullins-Herring (MH) universality class [15]) and is described, in Fourier components, by the equation

$$\frac{\partial h_q}{\partial t} = -Kq^4 h_q + \eta_q(t), \quad (1)$$

where $h_q$ is the Fourier component of the height measured relative to its spatial average, and $\eta_q(t)$ is the fluctuation of the rate of deposition, which is assumed to have a Gaussian distribution with zero mean and generally satisfies

$$\langle \eta_q(t)\eta_{q'}(t')\rangle = 2D_0 q^{-2\rho}\delta_{q,-q'}\delta(t-t'), \quad (2)$$

where $D_0$ is a constant, $\rho$ is a parameter that can be either positive or negative (actually in the case of conserved noise, which is of great interest in the MBE system, $\rho = -1$, see re. [16]), and $\delta_{q,-q'}$ is just the Kronecker symbol.

However, just like in the non-conservative case, a non-linear extension was needed to describe the richness of the MBE processes. Various symmetry arguments, originally suggested by Villain [4, 20] as well as some physical arguments [1] indicate that the relevant MBE growth equation is

$$\frac{\partial h_q}{\partial t} = -Kq^4 h_q - \frac{g}{\sqrt{\Omega}}\sum_{\ell,m} q^2 (\vec{\ell}\cdot\vec{m})\delta_{q,\ell+m} h_\ell h_m + \eta_q, \quad (3)$$

where g is the coupling constant, $\Omega$ is the volume of the system, to be taken eventually to infinity, and $\eta_q$ is the noise term. This equation, which is by no means trivial, has been analyzed later using the Dynamical Renormalization Group method (DRG) [6]. The theoretical predictions for the critical exponents made, using this method, agreed quite well with results of numerical integration of eq. (3) as well as with results of simulations of discrete models belonging to the MBE universality class (see for example [1], [15] and references therein). Therefore, these results became widely accepted in the community of surface-growth physicists.

However, in recent years some researchers raised again the question of the validity of the DRG theoretical predictions. One line of criticism was taken by DasSarma [15] who pointed out that the DRG results are derived from a leading order $\varepsilon$ – expansion of a one-loop renormalization analysis, where $\varepsilon = 4 - d$ (d being the substrate dimension). He stressed the point that for the relevant dimensions discussed in the literature, i.e. $d = 1$ or $2$, the expansion parameter $\varepsilon = 3$ or $2$ is not small, therefore one may legitimately question the validity of the calculated exponents.

A somewhat different line of criticism, however more radical and explicit was taken by Janssen [16] who was able to show that a two-loop calculation gives non trivial (although small) corrections to the critical exponents predicted by one-loop DRG calculation (more specifically $\alpha_{two-loop} = \alpha_{one-loop} - \delta$, where $\alpha$ is the roughness exponent and $\delta$ is the small correction). By doing so he actually made a substantial contribution in refuting the underlying assumption that the coupling constant renormalizes trivially. This assumption was very essential to the one-loop DRG calculations done so far. Janssen was also able to show explicitly the reason for this discrepancy, which has to do with a mathematically ill defined generalization of so called the Galilean invariance (actually tilt invariance) of the KPZ equation [3] suggested by Sun, Guo & Grant [17]. As mentioned above, Janssen found that the correction to the scaling exponents was very small, and he suspected that the smallness of the correction was related to many, but incomplete, cancellations between diagrams as well as within internal momentum integrals. On that basis he speculated that a mode-coupling approach is a useful approach for this problem.

In this paper I apply a method developed by Schwartz and Edwards [11-13] (also known as the Self-Consistent-Expansion (SCE) approach). This method has been previously applied to the KPZ equation. The method gained much credit in being able to give a sensible prediction for the KPZ critical exponents in the strong coupling phase, while DRG was not able to give any prediction for that phase for $d > 1$ (only the weak coupling solution was addressed). It is worth mentioning that this method is closely related to the mode-coupling approaches, in the sense that similar (but not identical) equations are obtained, while the underlying derivation is different. It is therefore, hoped that this paper will help to decide this unresolved situation, thus facing the challenge set up by Janssen [16]. I obtain the original results of the one-loop calculation [6, 8], thus corroborating these results, while avoiding the mathematical pathologies faced by the DRG method. This situation where DRG results obtained from different orders in $\varepsilon$ give different conceptual scenarios (i.e. trivial Vs. nontrivial renormalization of the coupling constant) calls for a resolution.

Another remarkable advantage of the SCE method is the minor changes needed in order to generalize the result with uncorrelated noise to include noise correlated in space. The above implies a second important motivation for this paper, namely a demonstration of the robustness of the SCE method as well as its mathematical coherence and consistency.

The SCE method is based on going over from the Fourier transform of the MBE equation in Langevin form to a Fokker-Planck form and constructing a self-consistent expansion of the distribution of the field concerned. The expansion is formulated in terms of $\phi_q$ and $\omega_q$, where $\phi_q$ is the two-point function in momentum space, defined by $\phi_q = \langle h_q h_{-q} \rangle_S$, (the subscript S denotes steady state averaging), and $\omega_q$ is the characteristic frequency associated with $h_q$, defined by $\omega_q^{-1} \equiv \int_0^\infty \langle h_q(t) h_{-q}(0) \rangle dt \Big/ \phi_q$.

I expect that for small enough q, $\phi_q$ and $\omega_q$ are power laws in q,

$$\phi_q = Aq^{-\Gamma} \qquad \text{and} \qquad \omega_q = Bq^\mu \qquad (4)$$

[Since dynamic surface growth is a remarkably multidisciplinary field, there are almost as many notations as there are workers in the field. Therefore I give a brief translation of our notations to those most frequently used:

$$\mu = z, \qquad \alpha = (\Gamma - d)/2, \qquad \text{and} \qquad \beta = \alpha/z = (\Gamma - d)/2\mu. \qquad (5)]$$

The method produces, to second order in this expansion, two nonlinear coupled integral equations in $\phi_q$ and $\omega_q$, that can be solved exactly in the asymptotic small q limit to yield the required scaling exponents governing the steady state behavior and the time evolution.

I begin with writing the Fokker-Planck form of the MBE equation (eq. (3))

$$\frac{\partial P}{\partial t} + \sum_q \frac{\partial}{\partial h_q}\left[D_{0q}\frac{\partial}{\partial h_{-q}} + K_q h_q + \sum_{\ell,m} M_{q\ell m} h_\ell h_m\right]P = 0, \qquad (6)$$

where $K_q = Kq^4$, $D_{0q} = q^{-2\rho}$ and $M_{q\ell m} = \frac{g}{\sqrt{\Omega}}q^2(\vec{\ell}\cdot\vec{m})\delta_{q,\ell+m}$.

A self-consistent expansion for such an equation was derived in the past (ref. [11-13]). The main idea is to write the Fokker-Planck equation $\partial P/\partial t = OP$ in the form $\partial P/\partial t = [O_0 + O_1 + O_2]P$, where $O_0$ is to be considered zero order in some parameter $\lambda$, $O_1$ is first order and $O_2$ is second order. The evolution operator $O_0$ is chosen to have a simple form $O_0 = -\sum_q \frac{\partial}{\partial h_q}\left[D_q \frac{\partial}{\partial h_{-q}} + \omega_q h_q\right]$, where $D_q/\omega_q = \phi_q$. Note that at present $\phi_q$ and $\omega_q$ are not known. I obtain next an equation for the two-point function. The expansion has the form $\phi_q = \phi_q + c_q\{\phi_p, \omega_p\}$, because the lowest order in the expansion already yields the unknown $\phi_q$. In the same way an expansion for $\omega_q$ is also obtained in the form $\omega_q = \omega_q + d_q\{\phi_p, \omega_p\}$. Now, the two-point function and the characteristic frequency are thus determined by the two coupled equations

$$c_q\{\phi_p,\omega_p\}=0 \quad \text{and} \quad d_q\{\phi_p,\omega_p\}=0. \qquad (7)$$

Working to second order in the expansion, one gets the two coupled integral equations

$$D_{0q}-K_q\phi_q+2\sum_{\ell,m}\frac{M_{q\ell m}M_{q\ell m}\phi_\ell\phi_m}{\omega_q+\omega_\ell+\omega_m}-2\sum_{\ell,m}\frac{M_{q\ell m}M_{\ell m q}\phi_m\phi_q}{\omega_q+\omega_\ell+\omega_m}-2\sum_{\ell,m}\frac{M_{q\ell m}M_{m\ell q}\phi_\ell\phi_q}{\omega_q+\omega_\ell+\omega_m}=0, \quad (8)$$

$$K_q-\omega_q-2\sum_{\ell,m}M_{q\ell m}\frac{M_{\ell m q}\phi_m+M_{m\ell q}\phi_\ell}{\omega_\ell+\omega_m}=0, \qquad (9)$$

where in deriving the last equation I have used the Herring consistency equation [14]. In fact Herring's definition of $\omega_q$ is one of many possibilities, each leading to a different consistency equation. But it can be shown, as previously done in [12], that this does not affect the exponents (universality).

In the following I will treat the equations (8) and (9) for our specific problem of interest (i.e. use the specific form of $K_q$ and $M_{qlm}$ for MBE). These equations can be rewritten as

$$D_0 q^{-2\rho}-Kq^4\phi_q+I_1(q)\phi_q+I_2(q)=0 \qquad (10)$$

$$\omega_q-Kq^4+J(q)=0, \qquad (11)$$

where the functions $I_1(q)$, $I_2(q)$ and $J(q)$ are given by

$$I_1(q)=-\frac{2g^2}{(2\pi)^d}\int d^d\ell\,\frac{q^2\left[\vec{\ell}\cdot(\vec{q}-\vec{\ell})\right]}{\omega_q+\omega_\ell+\omega_{q-\ell}}\left[\left|\vec{q}-\vec{\ell}\right|^2\vec{\ell}\cdot\vec{q}\phi_\ell+\ell^2(\vec{q}-\vec{\ell})\cdot\vec{q}\phi_{q-\ell}\right], \quad (12)$$

$$I_2(q)=\frac{2g^2}{(2\pi)^d}\int d^d\ell\,\frac{q^4\left[\vec{\ell}\cdot(\vec{q}-\vec{\ell})\right]^2}{\omega_q+\omega_\ell+\omega_{q-\ell}}\phi_\ell\phi_{q-\ell}, \qquad (13)$$

$$J(q)=-\frac{2g^2}{(2\pi)^d}\int d^d\ell\,\frac{q^2\left[\vec{\ell}\cdot(\vec{q}-\vec{\ell})\right]}{\omega_\ell+\omega_{q-\ell}}\left[\left|\vec{q}-\vec{\ell}\right|^2\vec{\ell}\cdot\vec{q}\phi_\ell+\ell^2(\vec{q}-\vec{\ell})\cdot\vec{q}\phi_{q-\ell}\right]. \quad (14)$$

As previously stated (eq. (4)) I expect that for small enough q, $\phi_q$ and $\omega_q$ are power laws in q, for small q (i.e. $\phi_q = Aq^{-\Gamma}$ and $\omega_q = Bq^\mu$). I am interested in eqs. (10) and (11) for small q's only. But, in order to achieve that one must consider the contribution of the large $\vec{\ell}$ integration on the small q behavior of the whole integrals. So I break up the integrals $I_i(q)$ and $J(q)$ into the sum of two contributions $I_i^>(q)$, $J^>(q)$ and $I_i^<(q)$, $J^<(q)$, corresponding to domains of $\vec{\ell}$ integration, with high and low momentum respectively. I expand $I_i^>(q)$ and $J^>(q)$ for small q's and obtain the leading small-q behavior of the integrals, and after retaining only the leading terms, eqs. (10) and (11) reduce now to

$$D_0 q^{-2\rho} + A_2 q^4 - Kq^4 \phi_q + A_1 q^6 \phi_q + I_1^<(q)\phi_q + I_2^<(q) = 0 \qquad (15)$$

$$\omega_q - Kq^4 + J^<(q) + A_3 q^6 = 0. \qquad (16)$$

At the mere price of renormalizing some constants in both equations, I am left with the integrals $I_1^<(q)$, $I_2^<(q)$ and $J^<(q)$ that can be calculated explicitly for small q's since for small $|\vec{\ell}|$'s the power law form for $\phi_\ell$ and $\omega_\ell$ (or $\phi_{q-\ell}$ and $\omega_{q-\ell}$) can be used (eq. (4)). In addition, the small-q dependence of each of the integrals naturally depends on the convergence of the integrals without cutoff. So, to leading order in q

$$I_1^<(q), J^<(q) \propto \begin{cases} q^6 & \text{for } d+4-\Gamma-\mu > 0 \\ q^{d+8-\Gamma-\mu} & \text{for } d+4-\Gamma-\mu < 0 \end{cases} \qquad (17)$$

$$I_2^<(q) \propto \begin{cases} q^4 & \text{for } d+4-2\Gamma-\mu > 0 \\ q^{d+8-2\Gamma-\mu} & \text{for } d+4-2\Gamma-\mu < 0 \end{cases} \qquad (18)$$

Going through the steps of a detailed analysis (like in the appendix of ref. [12]) I find that above $d_c = 4 + 2\rho$ a weak-coupling solution, where the exponents $\Gamma = 4 + 2\rho$ and $\mu = 4$ are obtained. The lower critical dimension of the non-linear MBE equation is thus $4 + 2\rho$.

A strong coupling solution can be obtained provided the $\Gamma$ and $\mu$ obey $d + 4 - \Gamma - \mu < 0$ and $d + 4 - 2\Gamma - \mu < 0$. In that case eqs. (15) and (16) take the form

$$D_0 q^{-2\rho} + A_2 q^4 - KA q^{4-\Gamma} + AA_1 q^{6-\Gamma} + \frac{2\lambda^2}{(2\pi)^d} \frac{A^2}{B} q^{d+8-2\Gamma-\mu} F(\Gamma,\mu) = 0, \quad (19)$$

$$B q^\mu - K q^4 + A_3 q^6 + \frac{2\lambda^2}{(2\pi)^d} \frac{A}{B} q^{d+8-\Gamma-\mu} G(\Gamma,\mu) = 0. \quad (20)$$

where $F(\Gamma,\mu)$ is given by

$$F(\Gamma,\mu) = -\int d^d t \frac{\vec{t} \cdot (\hat{e} - \vec{t})}{t^\mu + |\hat{e} - \vec{t}|^\mu + 1} \left[ |\hat{e} - \vec{t}|^2 (\vec{t} \cdot \hat{e}) t^{-\Gamma} + t^2 (\hat{e} - \vec{t}) \cdot \hat{e} |\hat{e} - \vec{t}|^{-\Gamma} \right] +$$
$$+ \int d^d t \frac{[\vec{t} \cdot (\hat{e} - \vec{t})]^2}{t^\mu + |\hat{e} - \vec{t}|^\mu + 1} t^{-\Gamma} |\hat{e} - \vec{t}|^{-\Gamma} \quad (21)$$

and $G(\Gamma,\mu)$ is given by

$$G(\Gamma,\mu) = -\int d^d t \frac{\vec{t} \cdot (\hat{e} - \vec{t})}{t^\mu + |\hat{e} - \vec{t}|^\mu} \left[ |\hat{e} - \vec{t}|^2 (\vec{t} \cdot \hat{e}) t^{-\Gamma} + t^2 (\hat{e} - \vec{t}) \cdot \hat{e} |\hat{e} - \vec{t}|^{-\Gamma} \right]. \quad (22)$$

From the conditions given above for a strong-coupling solution I find $d + 8 - \Gamma - \mu < 4$. Therefore, the last term in eq. (20) is dominant over the second term. Two possibilities seem to arise now. Either the last term dominates the first term in eq. (20), which implies $G(\Gamma,\mu) = 0$, or these terms are proportional to the same power of q, which implies the scaling relation $d + 8 - \Gamma - 2\mu = 0$. The first possibility requires $d + 8 - \Gamma - 2\mu < 0$, which is inconsistent with the whole idea of the expansion. The point is that higher order corrections have additional powers of $q^{d+8-\Gamma-2\mu}$ so that in our case the requirement $d + 8 - \Gamma - 2\mu < 0$ means that higher order terms are more violent than lower order ones (for small q's). Such a situation either implies inconsistency of the expansion, or calls for summing up the whole series in order to get a meaningful result. I assume that the expansion is consistent so that I am left with the second possibility, i.e. $d + 8 - \Gamma - 2\mu = 0$. It is interesting to mention here that Janssen's result is also consistent with the assumption that the requirement $d + 8 - \Gamma - 2\mu < 0$ cannot be fulfilled.

As for eq. (19), I get $d + 8 - 2\Gamma - \mu < 4 - \Gamma$, meaning that the third term is negligible compared to the last. Here again, I am faced with two possibilities. Either the first term and the last term are proportional to the same power of q, resulting in $d + 8 - 2\Gamma - \mu = -2\rho$, or the last term is dominant over the first term, in which case $d + 8 - 2\Gamma - \mu < -2\rho$, and $F(\Gamma, \mu) = 0$. A careful numerical calculation shows that $F(\Gamma, \mu) \neq 0$ for $d \leq 4$, so that the second possibility is ruled out. Therefore, I am left with the first possibility, namely with $d + 8 - 2\Gamma - \mu = -2\rho$.

I am led to the conclusion that the strong coupling solution is $\Gamma = (d + 4\rho + 8)/3$ and $\mu = (d - 2\rho + 8)/3$. Taking into account the condition for a strong-coupling solution given above, I find that such a solution is valid only for $d < d_c = 4 + 2\rho$. For $d > d_c$ the exponents describing the system are the same as those of the linear MBE equation, i.e. $\Gamma = d$ and $\mu = 4$. It should be mentioned that The $\rho$-dependent strong coupling solutions exist as long as $\rho \geq (d-4)/2$ (for lower values of $\rho$ we are actually always below the critical dimension, and the critical exponents are consequently always $\Gamma = d$ and $\mu = 4$). The translation of these results to the frequently used notation reads

$$z = \begin{cases} (d - 2\rho + 8)/3 & d < 4 + 2\rho \\ 4 & d > 4 + 2\rho \end{cases}, \qquad (23)$$

$$\alpha = \begin{cases} (2\rho + 4 - d)/3 & d < 4 + 2\rho \\ 0 & d > 4 + 2\rho \end{cases}. \qquad (24)$$

The final conclusion is that the second order self-consistent expansion yields results that corroborate the results of one-loop DRG [6, 8]. As mentioned at the beginning of the paper, the mode coupling approach give similar equations to those of SCE to second order - although a different derivation and analysis (i.e. summation of the perturbation series while neglecting vertex renormalization Vs. a perturbation theory for the Fokker-Planck form). On that basis, I expect the same results from a mode-coupling approach when applied to this problem.

To evaluate these results two facts have to be taken into account. The first is that the SCE approach does not rely on the symmetry argument [17] that its possible weakness was pointed out by Janssen. The second is that the SCE is known in other cases (i.e KPZ, see refs. [11-13]) to yield results that deviate from the results of the one-loop DRG and agree much better with simulations. This suggests that the one-loop DRG results may be exact for the case of MBE. On the other hand, attempts to verify Janssen's results via numerical simulations indeed found such corrections [18-19]. However, the corrections were systematically much larger than those predicted by Janssen himself. For example, for $d=1$ the deviation $\delta$ from the one-loop result differs from Janssen's result by an order of magnitude (correction of $\delta = 0.0025$ Vs. $\delta = 0.02$). Since $\delta = 0.0025$ is closer to $\delta = 0$ than to $\delta = 0.02$, this suggests that the deviation found in the numerical simulations might be related to some other factors. Clearly further investigation is needed to decide this interesting issue.